\documentclass{nature}
\usepackage{xcolor}
\usepackage{graphicx}
\usepackage{url}
\usepackage{amssymb,amsmath}
\usepackage{lineno}
\usepackage{bm}

\iftrue
\makeatletter
\let\saved@includegraphics\includegraphics
\AtBeginDocument{\let\includegraphics\saved@includegraphics}
\renewenvironment*{figure}{\@float{figure}}{\end@float}
\makeatother
\fi

\title{Extended-range statistical ENSO prediction through operator-theoretic techniques for nonlinear dynamics}

\author{Xinyang Wang$^{1}$, Joanna Slawinska$^{2}$ \& Dimitrios Giannakis$^{1}$}

\begin{document}

\maketitle

\begin{affiliations}
\item Center for Atmosphere Ocean Science, Courant Institute of Mathematical Sciences, New York University, New York, New York, USA
\item Department of Physics, University of Wisconsin-Milwaukee, Milwaukee, Wisconsin, USA

\end{affiliations}


\begin{abstract}
    Forecasting the El Ni\~no-Southern Oscillation (ENSO) has been a subject of vigorous research due to the important role of the phenomenon in climate dynamics and its worldwide socioeconomic impacts. Over the past decades, numerous models for ENSO prediction have been developed, among which statistical models approximating ENSO evolution by linear dynamics have received significant attention owing to their simplicity and comparable forecast skill to first-principles models at short lead times. Yet, due to highly nonlinear and chaotic dynamics (particularly during ENSO initiation), such models have limited skill for longer-term forecasts beyond half a year. To resolve this limitation, here we employ a new nonparametric statistical approach based on analog forecasting, called kernel analog forecasting (KAF), which avoids assumptions on the underlying dynamics through the use of nonlinear kernel methods for machine learning and dimension reduction of high-dimensional datasets. Through a rigorous connection with Koopman operator theory for dynamical systems, KAF yields statistically optimal predictions of future ENSO states as conditional expectations, given noisy and potentially incomplete data at forecast initialization.  Here, using industrial-era Indo-Pacific sea surface temperature (SST) as training data, the method is shown to successfully predict the Ni\~no~3.4 index in a 1998--2017 verification period out to a 10-month lead, which corresponds to an increase of 3--8 months (depending on the decade) over a  benchmark linear inverse model (LIM), while significantly improving upon the ENSO predictability ``spring barrier''. In particular, KAF successfully predicts the historic 2015/16 El Ni\~no at initialization times as early as June 2015, which is comparable to the skill of current dynamical models. An analysis of a 1300-yr control integration of a comprehensive climate model (CCSM4) further demonstrates that the enhanced predictability afforded by KAF holds over potentially much longer leads, extending to 24 months versus 18 months in the benchmark LIM. Probabilistic forecasts for the occurrence of El Ni\~no/La Ni\~na events are also performed, and assessed via information-theoretic metrics, showing an improvement of skill over LIM approaches, thus opening an avenue for environmental risk assessment relevant in a variety of contexts. 


\end{abstract}

Previous studies on improving the skill of conventional LIMs\cite{penland1993,penland1995} have highlighted the importance of nonlinearity in ENSO dynamics, such as surface-subsurface interactions and surface winds\cite{chapman2015}, which are usually underestimated in linear dynamics approximations. Adequate representation of such processes within statistical ENSO models is important to attain optimal long-term forecast skill, influencing two major components of model design, namely (i) the construction of predictor vectors to extract pertinent information about the state of the climate system; and (ii) the assumed evolution dynamics employed to make predictions beyond the training period. For instance, LIMs use the leading principal components from empirical orthogonal function (EOF) analysis as predictors, and approximate their tendencies by a Markov prediction model similar to a linear regression. A linear structure is imposed in two aspects here; that is, the linear predictors obtained by EOF analysis and the linear dynamical model treating these predictors as state vectors. This suggests that two corresponding types of improvement of conventional LIMs can be sought. Indeed, a number of studies\cite{kondrashov2005, chapman2015} have shown that inclusion of LIM residuals by the present and recent past states can implicitly capture subsurface forcing and its nonlinear interactions with SST, consequently contributing to more skillful ENSO predictions. On the other hand, replacing EOF analysis by nonlinear dimension reduction techniques\cite{lima2009} has also led to improved performance by LIMs under certain circumstances, especially for long-term ENSO forecasts. 

Other statistical approaches for ENSO prediction\cite{VanDenDool06,DingEtAl18,DingEtAl19,HamEtAl19} have included methods based on Lorenz's analog forecasting technique\cite{lorenz1969} or neural networks (NNs)\cite{LeCunEtAl15}. The core idea of analog methods\cite{VanDenDool06,DingEtAl18,DingEtAl19} is to employ observational or free-running (non-initialized) model data as libraries of states, through which forecasts can be made by identifying states in the library that closely resemble the observed data at forecast initialization, and following the evolution of the quantity of interest on these so-called analogs to yield a prediction. Advantages of analog forecasting include avoiding the need for an expensive, and potentially unstable, initialization system, as well as reducing structural model error if natural analogs are employed. For phenomena such as ENSO exhibiting a reasonably small number of effective degrees of freedom, the performance of analog techniques has been shown to be comparable to, and sometimes exceed, the skill of initialized forecasts by coupled general circulation models (CGCMs)\cite{DingEtAl18,DingEtAl19}. Meanwhile, NN models build a representation of the predictand as a nonlinear function of the predictor variables by optimization within a high-dimensional parametric class of functions. A recent study\cite{HamEtAl19} utilizing convolutional NNs trained on GCM output and reanalysis data has demonstrated forecast  skill for the 3-month averaged Ni\~no~3.4 index extending out to 17 months.

The KAF approach\cite{zhao2016,AlexanderGiannakis19} employed in this work can be viewed as a generalization of conventional analog forecasting, utilizing nonlinear-kernel algorithms for statistical learning\cite{CuckerSmale01} and operator-theoretic ergodic theory\cite{EisnerEtAl15,BudisicEtAl12} to optimally capture the evolution of a response variable (here, an ENSO index) under partially observed nonlinear dynamics. In particular, unlike conventional approaches, the output of KAF is not given by a local average over analog states (or linear combinations of states in the case of constructed analogs\cite{VanDenDool06}), but is rather determined by a projection operation onto a function space of observables of high regularity, learned from high-dimensional training data. This hypothesis space has as a basis a set of eigenvectors of a judiciously constructed kernel matrix, which depends nonlinearly on the input data, thus potentially capturing richer patterns of variability than the linear principal components from EOF analysis. It should be noted, in particular, that the eigenvectors employed in KAF are not expressible as linear projections of the data onto corresponding EOFs. By virtue of this structure, the forecast function from KAF can be shown to converge in an asymptotic limit of large data to the conditional expectation of the response function, acted upon by the Koopman evolution operator of the dynamical system over the desired lead time, conditioned on the input (predictor) data at forecast initialization\cite{AlexanderGiannakis19}. As is well known from statistics, the conditional expectation is optimal in the sense of minimal expected square error ($L^2$ error) among all forecasts utilizing the same predictors. 

Here, we build KAF models using lagged histories of Indo-Pacific SST fields as input data. Specifically, we employ the class of kernels introduced in the context of nonlinear Laplacian spectral analysis (NLSA)\cite{GiannakisMajda12a, GiannakisMajda12b} algorithms, whose eigenfunctions provably capture modes of high dynamical coherence\cite{Giannakis19,DasGiannakis19}, including representations of ENSO, ENSO combination modes, and Pacific decal variability\cite{SlawinskaGiannakis17,GiannakisSlawinska18,WangEtAl19}. This capability is realized without ad hoc prefiltering of the input data  through the use of delay-coordinate maps\cite{SauerEtAl91} and a nonlinear (normalized Gaussian) kernel function\cite{coifman2006diffusion} measuring similarity between delay-embedded sequences of SST fields.  A key advantage of the methodology employed here over LIMs or other parametric models is its nonparametric nature, in the sense that KAF does not impose explicit assumptions on the dynamical model structure (e.g., linearity), thus avoiding systematic model errors that oftentimes preclude parametric models from attaining useful skill at long lead times. Moreover, by employing delays, KAF enhances the information content of the predictor data beyond the information present in individual SST snapshots, allowing it to capture important contributors to ENSO predictability, such as upper-ocean heat content and surface winds\cite{KleemanEtAl95,chapman2015}. 

A distinguishing aspect of the approach presented here over recent analog\cite{DingEtAl18,DingEtAl19} and NN\cite{HamEtAl19} methods utilizing large, GCM-derived, multivariate datasets for training is that our observational forecasts are trained only on industrial-era observational/reanalysis SST fields. Indeed, due to the large number of parameters involved, NN methods cannot be adequately trained from industrial-era data alone\cite{HamEtAl19}, and are trained instead using thousands of years of model output. In contrast, KAF is an intrinsically nonparametric method, involving only a handful of meta-parameters (such as embedding window and hypothesis space dimension), and thus able to yield robust ENSO forecasts from a comparatively modest training dataset spanning $ \sim 120 $ years. It should also be noted that KAF exhibits theoretical convergence and optimality guarantees\cite{AlexanderGiannakis19} which, to our knowledge, are not available for other comparable statistical ENSO forecasting methodologies. 


\section*{Results}
We present prediction results obtained via KAF and LIMs in a suite of prediction experiments for (i) the Ni\~no 3.4 index in industrial-era HadISST data\cite{RaynerEtAl03}; (ii) the Ni\~no 3.4 index in a 1300-yr control integration of CCSM4\cite{GentEtAl11}, a CGCM known to exhibit fairly realistic ENSO dynamics and Pacific decadal variability\cite{DeserEtAl12b}; and (iii) the probability of occurrence of El Ni\~no/La Ni\~na events in both HadISST and CCSM4. All experiments use SST fields from the respective datasets in a fixed Indo-Pacific domain as training data. The Ni\~no~3.4 prediction skill is assessed using root-mean-square error (RMSE) and pattern correlation (PC) scores. We employ $ \text{PC} = 0.6 $ and 0.5 as representative thresholds separating useful from non-useful predictions. The probabilistic El Ni\~no/La Ni\~na experiments are assessed using the Brier skill score (BSS)\cite{Brier50,WeigelEtAl07}, as well as information-theoretic relative entropy metrics\cite{Kleeman02,DelSoleTippett07,GiannakisEtAl12b}. For consistency with an operational forecasting environment, all experiments employ disjoint, temporally consecutive portions of the available data for training and verification, respectively. In the CCSM4 experiments, Ni\~no~3.4 indices are computed relative to monthly climatology of the training data; in HadISST we use anomalies relative to the 1981--2010 monthly climatology as per current convention in many modeling centers\cite{LHeureuxEtAl17}.   

\begin{figure}
\centering
\includegraphics[width=\linewidth]{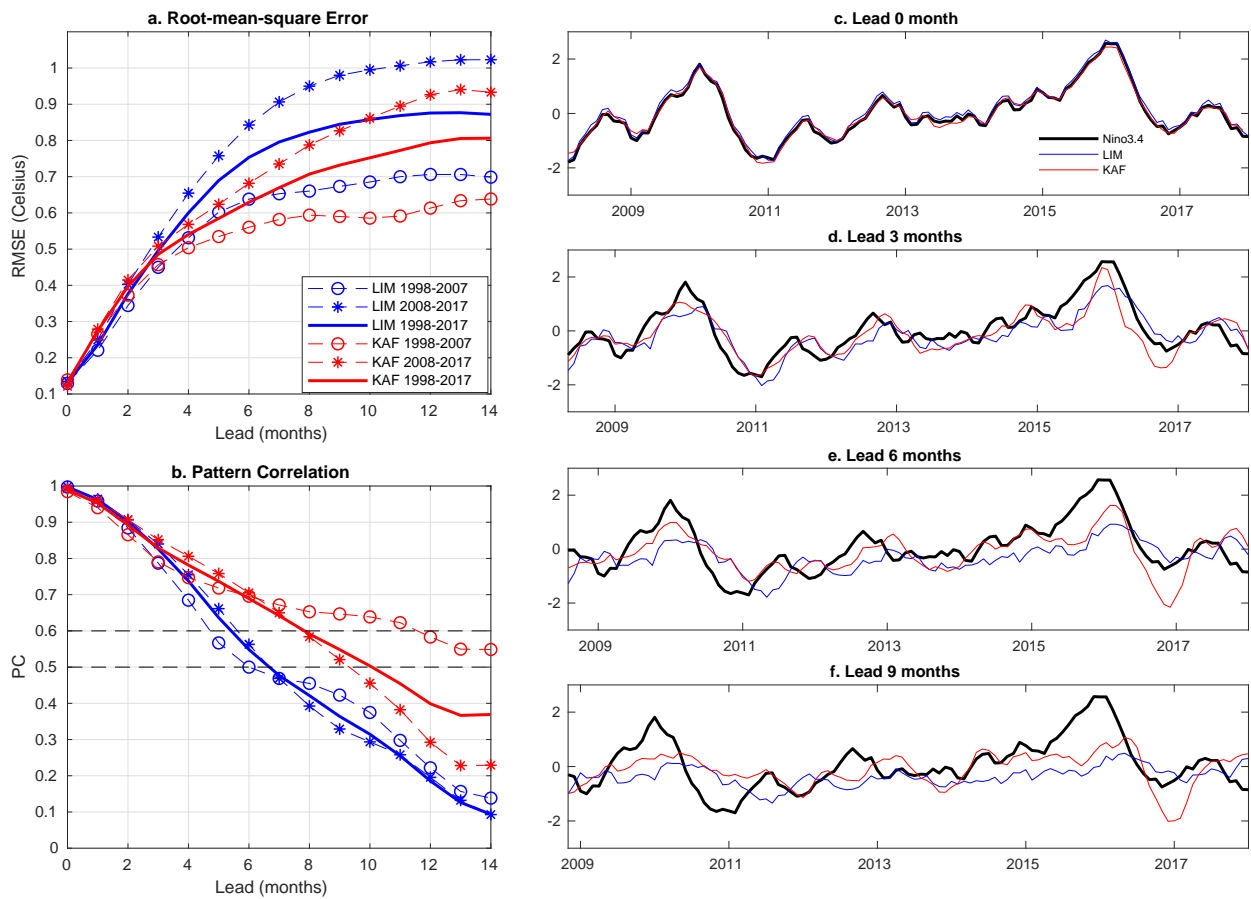}
\caption{Forecasts of the Ni\~no 3.4 index in industrial-era HadISST data during 1998--2017  obtained via KAF (red) and LIM (blue). (a, b) RMSE and PC scores, respectively, as a function of lead time, computed for 1998--2007 ($\circ$ markers), 2008--2017 ($*$ markers), and the full 1998--2017 verification period (solid lines). The 0.6 and 0.5 PC levels are highlighted in (b) for reference. (c--f) Running forecasts for representative lead times in the range 0--9 months. Black lines show the true signal at the corresponding verification times.}
\label{Fig1}
\end{figure}

Figure~\ref{Fig1} displays KAF and LIM prediction results  for the Ni\~no 3.4 index in the HadISST dataset over a 1998--2017 verification period, using 1870--1997 for training. In Figure~\ref{Fig1}(a,b) it is clearly evident that even though there is significant decadal variability of skill, KAF consistently outperforms the LIM in terms of both RMSE and PC scores, with an apparent increase of 4--7 months of useful forecast horizon. In particular, KAF attains the longest useful skill in the first decade of the verification period, 1998--2007 (despite an initially faster decrease of skill over 0--6 months), where the PC score is seen to cross the  0.6 threshold at approximately 12 months, and remains above 0.5 at least out to 14 months, compared to 5 and 6 months for LIM, respectively. In 2008--2017, the PC skill of KAF exhibits a more uniform rate of decay, crossing the $ \text{PC}= 0.5 $ threshold at lead time of 9 months, compared to 6--7 months for LIM. When measured over the full, 20-year verification period, KAF provides 3 to 4 months of additional useful skill, as well as a markedly slower rate of skill reduction beyond 4 months. As a comparison with state-of-art dynamical forecasts, the skill scores in Figure~\ref{Fig1} are generally comparable, and in some cases exceed, recently reported scores from the North American Multimodel Ensemble\cite{BarnstonEtAl19} (though it should be kept in mind that the verification periods of that study and ours are different). 

In separate calculations, we have verified that incorporating delays in the kernel is a significant contributor to the skill of KAF, particularly at lead times greater than 2 months. This is consistent with previous work, which has shown that delay embedding increases the capability of kernel algorithms to extract dynamically intrinsic coherent patterns in the spectrum of the Koopman operator\cite{Giannakis19,DasGiannakis19}, including ENSO and its linkages with seasonal and decadal variability of the climate system\cite{SlawinskaGiannakis17,GiannakisSlawinska18,WangEtAl19}. Given that our main focus here is on the extended-range regime, Figure~\ref{Fig1} depicts KAF results for a single embedding window of 12 months. Operationally, however, one would employ different embedding windows at different leads to optimize performance. 

\begin{figure}
    \centering
    \includegraphics[width=\linewidth]{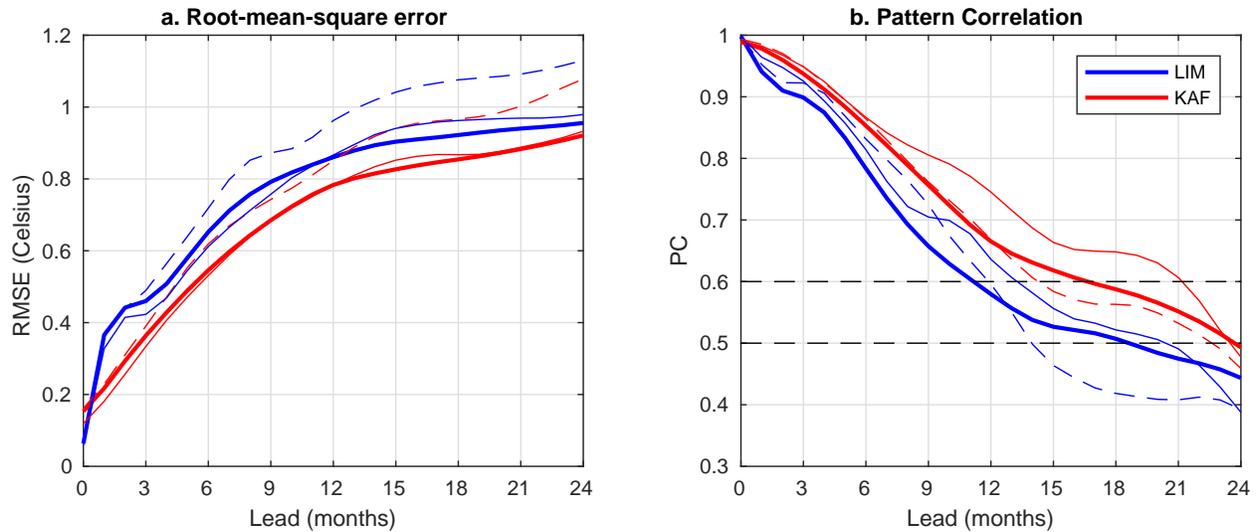}
    \caption{\label{figCCSM4Skill}As in Figure~1(a,b), but for a 200-year verification period corresponding to CCSM4 simulation years 1100--1300. Solid and dashed lines correspond to scores computed from the fourth and last decade of the verification period, respectively.}
\end{figure}

Next, in order to assess the potential skill of KAF and LIM approaches in an environment with more data available for training and verification (contributing to  more robust model construction and skill assessment, respectively) than HadISST, we examine Ni\~no 3.4 prediction results for the CCSM4 dataset, using the first 1100 years for training and the last 200 years for verification. Figure~\ref{figCCSM4Skill} demonstrates that while both KAF and LIM exhibit significantly higher skill compared to the HadISST results in Figure~\ref{Fig1}(a, b), the performance of KAF is particularly strong, with PC scores computed over the 200 year verification period crossing the 0.6 and 0.5 levels at 17- and 24-month leads, respectively. The latter values correspond to an increase of predictability over HadISST by 9 and 14 months, measured with respect to the same PC thresholds. In comparison, the $ \text{PC} = 0.6 $ and 0.5 predictability horizons for LIM are 11 and 18 months, respectively, corresponding to an increase of 6 and 11 months over HadISST. As illustrated in Figure~\ref{figCCSM4Skill}, the RMSE and PC scores from individual decades in the CCSM4 verification period exhibit significant variability, and similarly to HadISST, KAF consistently outperforms LIM on individual verification decades. In general, KAF appears to benefit from the larger CCSM4 dataset more than LIM, which is consistent with the former method better leveraging the information content of large data sets due to its nonparametric nature and theoretical ability to consistently approximate an optimal prediction function (the conditional expectation). In contrast, LIM's performance is limited to some extent by structural model error due to linear dynamics, and this type of error may not be possible to eliminate even with arbitrarily large training datasets.

\begin{figure}
\centering
\includegraphics[width=\linewidth]{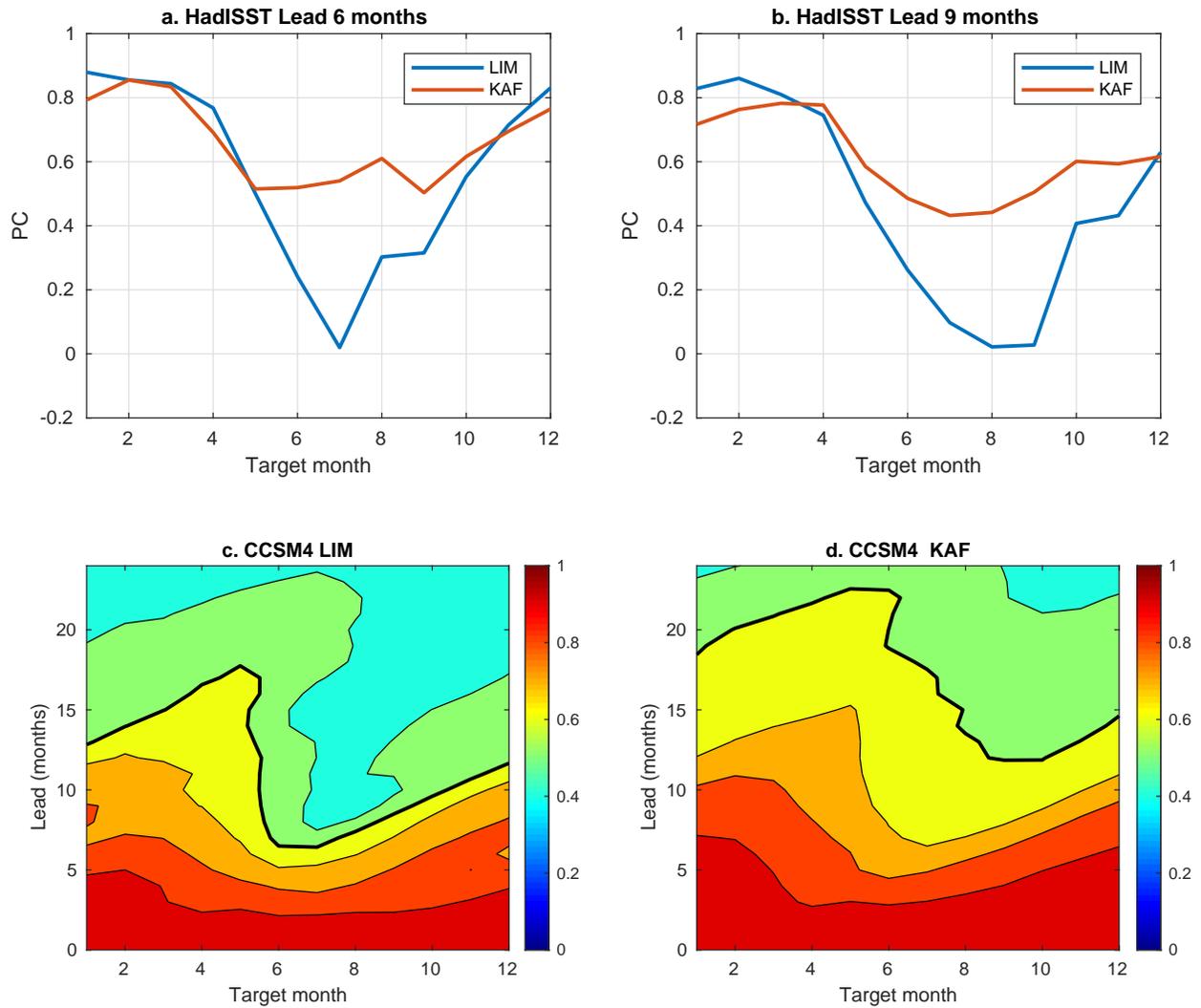}
\caption{(a, b) Seasonal dependence of Ni\~no~3.4 PC scores for KAF (red lines) and the benchmark LIM at 6- and 9-month leads in industrial-era HadISST data during 1998--2017. (c, d) LIM and KAF PC scores for 0--24-month leads in CCSM4 during simulation years 1100--1300. Contours are drawn at $ \text{PC} = 0.9, 0.8, \ldots, 0.5$. Thick black lines highlight  $\text{PC} = 0.6 $ contours for reference.}
\label{Fig2}
\end{figure}

Another important consideration in ENSO prediction is the seasonal dependence of skill, exhibiting the so-called ``spring barrier''\cite{BarnstonRopelewski91}. Figure~\ref{Fig2} shows the month-to-month distribution of the KAF and LIM Ni\~no~3.4 PC scores, computed for 6- and 9-month leads in HadISST and 0--24-month leads in CCSM4 using the same training and verification datasets as in Figures~\ref{Fig1} and~\ref{figCCSM4Skill}, respectively. These plots feature characteristic spring barrier curves, with the highest predictability occurring in late winter to early spring and a clear drop of skill in summer\cite{chapman2015,kondrashov2005}. The diminished summer predictability is thought to be caused by the low amplitude of SST anomalies developing then, making ENSO dynamics more sensitive to high-frequency atmospheric noise\cite{chapman2015}. As we have shown in previous work\cite{GiannakisSlawinska18}, the class of kernels employed here for forecasting is adept at capturing the effects of atmospheric noise and its nonlinear impact on ENSO statistics, including positive SST anomaly skewness underlying El Ni\~no/La Ni\~na asymmetry. The method has also been found highly effective in capturing the phase locking of ENSO to the seasonal cycle through a hierarchy of combination modes\cite{StueckerEtAl13,SlawinskaGiannakis17,WangEtAl19}. Correspondingly, while both methods exhibit a reduction of skill during summer, KAF fares significantly better than the benchmark LIMs in both of the HadISST and CCSM4 datasets. More specifically, in the HadISST results in Figure~\ref{Fig2}(a,b), the LIM's PC score drops rapidly to small, 0--0.3 values between June and September, while  KAF maintains scores of about 0.4 or greater over the same period. In particular, the June--September PC scores for KAF at 6-month leads hover between 0.5 and 0.6, which are values that one could consider as ``beating'' the spring barrier (albeit marginally).  Similarly, in CCSM4 (Figure~\ref{Fig2}(d)), the $\text{PC} = 0.6 $ contour for KAF decreases less appreciably and abruptly over May--August than in the case of the LIM, indicating that KAF is better at maintaining  skill during the transition from spring to summer. Since strong El Ni\~no events are often triggered by westerly wind bursts in spring, advecting warm surface water to the east\cite{chapman2015}, the better performance of KAF in summer suggests that it is more capable of capturing the triggering mechanism\cite{GiannakisSlawinska18}, aiding its ability in predicting the onset of ENSO. Indeed, this is consistent with the running forecast time series for HadISST in Figures~\ref{Fig1}(c,d), in which KAF often yields more accurate forecasts at the beginning of El Ni\~no events, e.g., during 2009/10 and 2015/16.

\begin{figure}
    \centering
    \includegraphics[width=.5\linewidth]{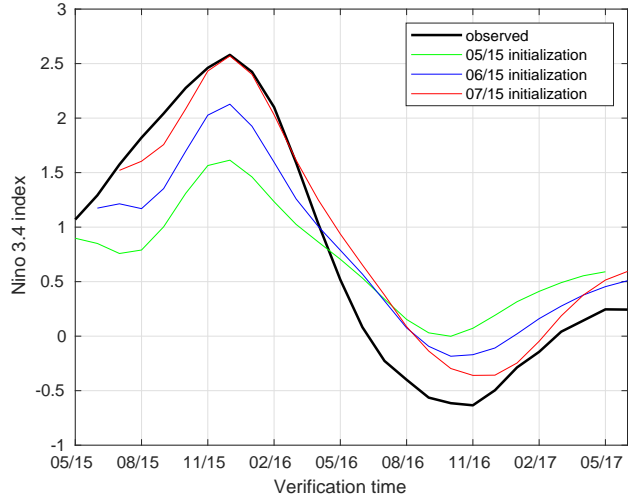}
    \caption{\label{figElNino15}KAF hindcasts of the 3-month-averaged Ni\~no~3.4 index in HadISST during the 2015/16 El Ni\~no period. Hindcast trajectories are initialized in the May, June, and July preceding the December 2015 peak of the event.}
\end{figure}

As a final deterministic hindcast experiment, we consider prediction of the 2015/16 El Ni\~no event with KAF. This event has received considerable attention by researchers and stakeholders across the world\cite{LHeureuxEtAl17}, in part due to it being the first major El Ni\~no since the 1997/98 event, providing an important benchmark to assess the advances in modeling and observing systems in the nearly two intervening decades. Figure~\ref{figElNino15} shows KAF forecast trajectories of 3-month running means of the Ni\~no~3.4 index in HadISST for the period May 2015 to May 2017, initialized in May, June, and July of 2015 (i.e., 7, 6, and 5 months prior to the December 2015 peak of the of the event). In this experiment, we use Indo-Pacific SST data from HadISST over the period January 1870 to December 2007 for training. Moreover, we perform prediction of 3-month-averaged Ni\~no 3.4 indices to highlight seasonal evolution\cite{LHeureuxEtAl17}. Prediction results for the instantaneous (monthly) Ni\~no 3.4 index are broadly consistent with the results in Figure~\ref{figElNino15}. 

We find that the KAF hindcast initialized in July 2015 accurately predicts the 3-month averaged Ni\~no~3.4 evolution through February 2016, including the 2015/16 El Ni\~no peak, yielding a moderate over-prediction over the ensuing months which nevertheless correlates well with the true signal. The trajectory initialized in June 2015 also predicts a strong ($\geq 2 $~$^\circ$C) El Ni\~no event peaking in December 2015, but underestimates the true 2.5~$^\circ$C anomaly peak by approximately 0.4~$^\circ $C. Meanwhile, the trajectory initialized in May 2015 predicts a 1.5~$^\circ $C positive anomaly in the ensuing December, which falls short of the true event by 1~$^\circ $C but still qualifies as an El Ni\~no event by conventional definitions. Overall, this level of performance compares favorably with many statistical model forecasts of the 2015/16 El Ni\~no (which typically did not yield a possibility of a $ \geq 2 $~$^\circ $C event until August 2015\cite{LHeureuxEtAl17}). In fact, the forecast trajectories in Figure~\ref{figElNino15} have comparable, and sometimes higher, skill than forecasts from many operational weather prediction systems, despite KAF utilizing far-lower numerical/observational resources and temporal resolution.

We now turn to the skill of probabilistic El Ni\~no/La Ni\~na forecasts. A distinguishing aspect of probabilistic prediction over point forecasts (e.g., the results in Figures~\ref{Fig1}--\ref{Fig2}) is that it provides more direct information about uncertainty, and as a result more actionable information to decision makers. Probabilistic prediction with either first-principles or statistical models, or ensembles of models, is typically conducted by binning collections of point forecasts, e.g., realized by random draws from distributions of initial conditions and/or model parameters. Here, we employ an alternative approach, which to our knowledge has not been previously pursued, whereby KAF and LIM approaches are used to predict the probability of occurrence of El Ni\~no or La Ni\~na events directly, without generating ensembles of trajectories. Our approach is based on the fact that predicting conditional probability is equivalent to predicting the conditional expectation of a characteristic function indexing the event of interest\cite{AlexanderGiannakis19}. As a result, this task can be carried out using the same KAF and LIM approaches described above, replacing the  Ni\~no 3.4 index by the appropriate characteristic function as the response variable. 

Here, we construct characteristic functions for El Ni\~no and La Ni\~na events in the HadISST and CCSM4 data using the standard criterion requiring that at least five overlapping seasonal (3-month running averaged) Ni\~no 3.4 indices to be greater (smaller) than 0.5 ($-0.5$) $^\circ$C, respectively\cite{LHeureuxEtAl17}. In this context, natural skill metrics are provided by the BSS\cite{Brier50,WeigelEtAl07}, as well as relative entropy from information theory\cite{Kleeman02,DelSoleTippett07} (see Methods). For binary outcomes, such as occurrence vs.\ non-occurrence of El Ni\~no or La Ni\~na events, the BSS is equivalent to a climatology-normalized RMSE score for the corresponding characteristic function. As information-theoretic metrics,  we employ two relative entropy functionals\cite{GiannakisEtAl12b}, denoted $\mathcal{D}$ and $\mathcal{E}$, which measure respectively the precision relative to climatology and ignorance (lack of accuracy) relative to the truth in a probabilistic forecast. For a binary outcome,  $\mathcal{D} $ attains a maximal value $\mathcal{D}_*$, corresponding to a maximally precise binary forecast relative to climatology. On the other hand, $\mathcal{E}$ is unbounded, but has a natural scale $\mathcal{E}_*$ corresponding to the ignorance of a probabilistic forecast based on random draws from the climatology; this makes $\mathcal{E} = \mathcal{E}_*$ a natural threshold indicating loss of skill.  Overall, a skillful binary probabilistic forecast should simultaneously have $ \text{BSS} \ll 1 $, $\mathcal{D} \simeq \mathcal{D}_*$, and $\mathcal{E} \ll \mathcal{E}_*$. Note that we only report values of these scores in the CCSM4 experiments, as we found that the HadISST verification is not sufficiently long for statistically robust computation of relative-entropy scores.   

\begin{figure}
\centering
\includegraphics[width=\linewidth]{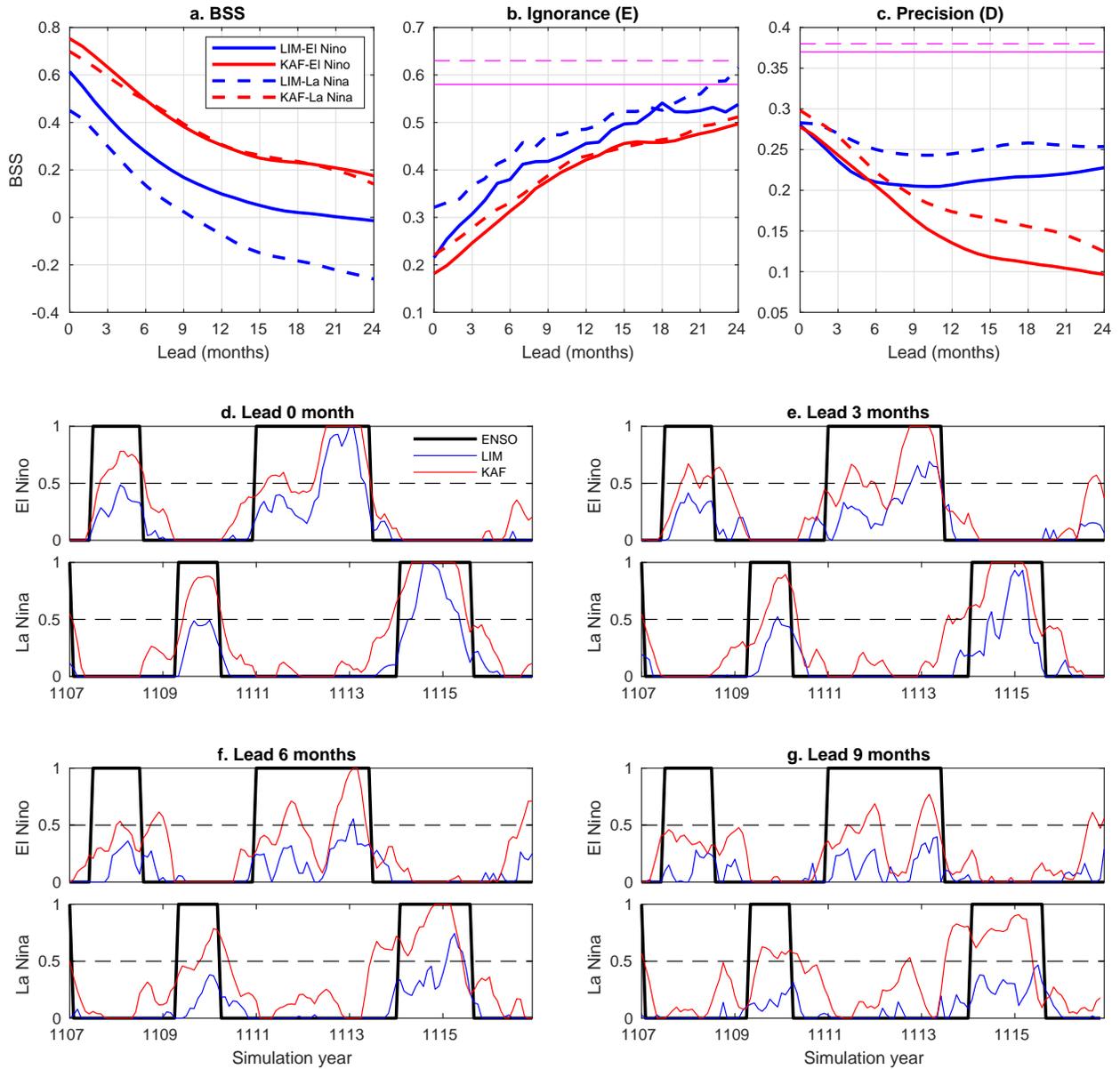}
\caption{Probabilistic El Ni\~no (solid lines) and La Ni\~na (dashed lines) forecasts in CCSM4 data during simulation years 1100--1300, using KAF (red lines) and LIMs (blue lines). (a--c) BSS and information-theoretic ignorance ($\mathcal{E}$) and precision ($\mathcal{D}$) metrics, respectively, as a function of lead time. Magenta lines in (b) and (c) indicate the entropy thresholds $\mathcal{E}_*$ and $\mathcal{D}_*$, respectively. (d--g) Running forecasts of the characteristic function for El Ni\~no/La Ni\~na events, representing conditional probability, for representative lead times in the range 0--9 months. Black lines show the true signal at the corresponding lead times.} 
\label{Fig3}
\end{figure}

\begin{figure}
\centering
\includegraphics[width=\linewidth]{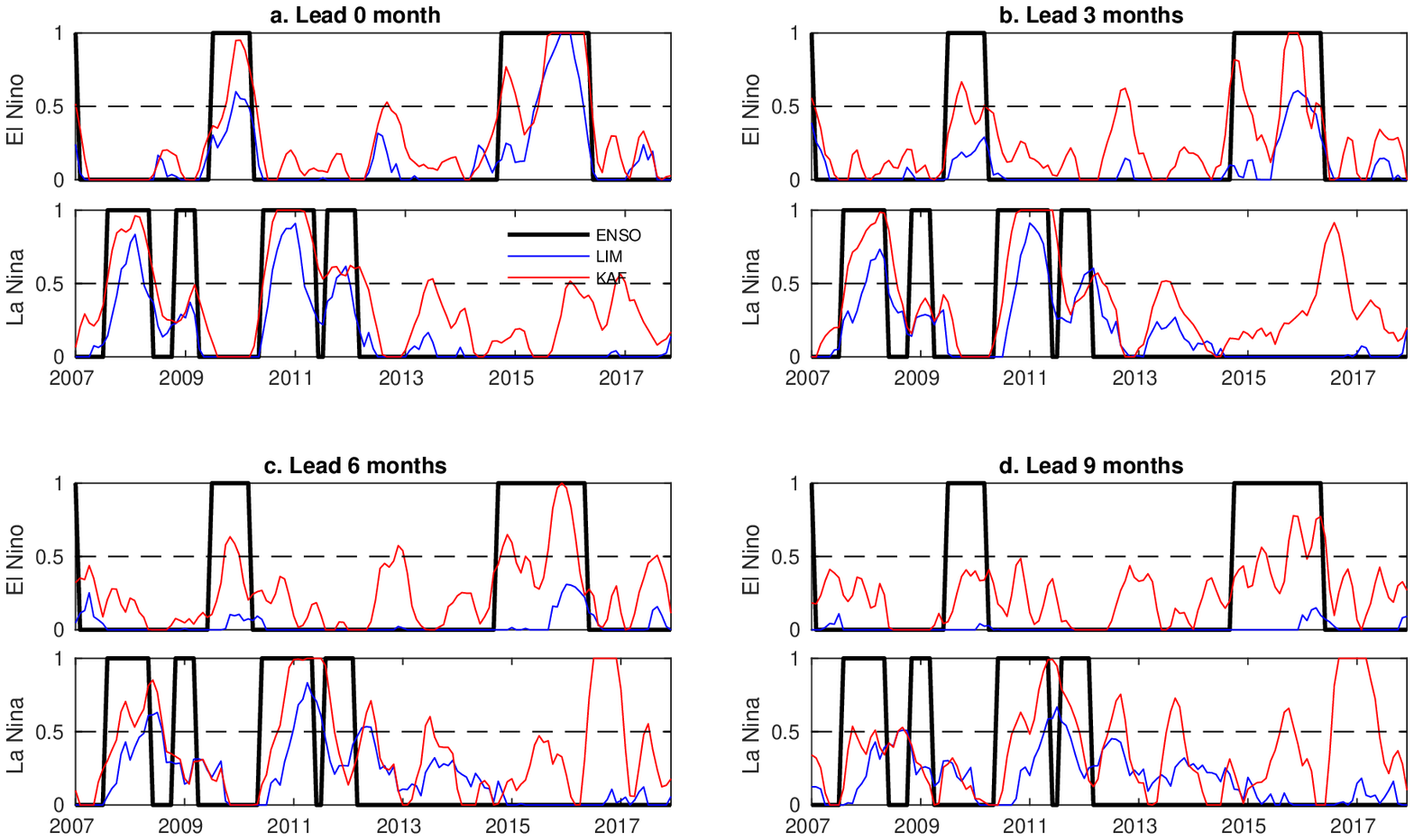}
\caption{As in Figure~\ref{Fig3}(d--g), but for probabilistic El Ni\~no/La Ni\~na forecasts in industrial-era HadISST data during 2007--2017.}
\label{Fig4}
\end{figure}

The results of probabilistic El Ni\~no/La Ni\~na prediction for CCSM4 and HadISST  are shown in Figures~\ref{Fig3} and Figure~\ref{Fig4}, respectively. As is evident in Figure~\ref{Fig3}(a, b), in CCSM4 KAF performs markedly better than LIM in terms of the BSS and $\mathcal{E}$ metric for all examined lead times, and for both El Ni\~no and La Ni\~na events. The $\mathcal{D}$ scores in Figure~\ref{Fig3}(c) start from $\simeq 0.3$ at forecast initialization (zero lead) for both methods, and while the KAF scores exhibit a monotonic decrease with lead time, in the case of LIM they exhibit an oscillatory behavior, hovering around $\simeq 0.25 $ values. The latter behavior, in conjunction with a steady decrease of BSS and $\mathcal{E}$ seen in Figure~\ref{Fig3}(a, b), is a manifestation of the fact that, as the lead time grows, LIM produces biases, likely due to dynamical model error. Note, in particular, that a forecast of simultaneously high ignorance (i.e., large error) and high precision, as the LIM forecast at late times, must necessarily be statistically biased as it underestimates uncertainty. In contrast, the KAF-derived results exhibit a simultaneous increase of ignorance and decrease of precision, as expected for an unbiased forecast under complex dynamics with intrinsic  predictability limits. Turning to the HadISST results, Figure~\ref{Fig4} again demonstrates that KAF performs noticeably better than LIM for both El Ni\~no and La Ni\~na prediction. Note that the apparent ``false positives'' in the KAF-based La Ni\~na results around 2017 are not necessarily unphysical. In particular, there are weak La Ni\~na events documented during 2016--2017 and 2017--2018, which are excluded from the true characteristic function for La Ni\~nas, but may exhibit residuals in the approximate characteristic function output by KAF.   

In conclusion, this work has demonstrated the efficacy of KAF in statistical ENSO prediction, with a robust improvement in useful prediction horizon in observational data by 3--7 months over LIM approaches, and three appealing characteristics---namely, more skillful forecasts at long lead times with slower reduction of skill, reduced spring predictability barrier, and improved prediction of event onset. Moreover, in the setting of model data with larger sample sizes, the enhanced performance of KAF becomes more pronounced, with skill extending out to 24 month leads. Aside from the higher skill in predicting ENSO indices, the method is also considerably more skillful in probabilistic El Ni\~no/La Ni\~na prediction. We attribute these improvements to the nonparametric nature of KAF, which can consistently approximate optimal prediction functions via conditional expectation in the presence of nonlinear dynamics through a rigorous connection with Koopman operator theory. A combination of this approach with delay-coordinate maps further aids its capability to extract dynamically coherent predictor variables (through kernel eigenfunctions), including seasonal variability associated with ENSO combination modes, representations of higher-order ENSO statistics due to  atmospheric noise, and Pacific decadal variability. Even though nonparametric ENSO prediction methods are not particularly common, this study shows that methods such KAF have high potential for skillful ENSO forecasts at long lead times, and can be naturally expected to be advantageous in forecasting other geophysical phenomena and their impacts, particularly in situations  where the underlying dynamics is unknown or partially understood.

\begin{methods}


\subsection{Datasets}
The observational data used in this study consists of monthly averaged SST fields from the  Hadley Centre Sea Ice and Sea Surface Temperature (HadISST) dataset\cite{RaynerEtAl03,HadISST}, sampled on a $1^{\circ} \times 1^{\circ}$ latitude--longitude grid, and spanning the industrial era, 1870--2017.  The modeled SST data are from a 1300-year, pre-industrial control integration of the Community Climate System Model version 4 (CCSM4)\cite{GentEtAl11,CCSM4}, monthly averaged, and sampled on the model's native ocean grid of approximately $1^{\circ}$ nominal resolution.  All experiments use SST on the Indo-Pacific longitude-latitude box $28^{\circ}$E--$70^{\circ}$W, $30^{\circ}$S--$20^{\circ}$N as input predictor data, and the corresponding Ni\~no 3.4 indices as target (predictand) variables. The latter are defined as SST anomalies relative to a monthly climatology (to be specified below) computed from each dataset, spatially averaged over the region $5^{\circ}$N--$5^{\circ}$S, $170^{\circ}$--$120^{\circ}$W. The number $p $ of spatial gridpoints in the HadISST and CCSM4 Indo-Pacific domains is 11,315 and 31,984, respectively.  

Each dataset is divided into disjoint, temporally consecutive, training and verification periods. In the HadISST results in Figures~\ref{Fig1}, \ref{Fig2}, and~\ref{Fig4}, these periods are January 1870 to December 1997 and January 1998 to December 2017, respectively. The 2015/16 El Ni\~no forecasts in Figure~\ref{figElNino15} use January 1870 to December 2014 for training and May 2015 to May 2017 for verification. In the CCSM4 results in Figures~\ref{figCCSM4Skill} and~\ref{Fig3}, the training and verification periods correspond to simulation years 1--1100 and 1101--1300, respectively. In separate calculations, we have removed portions of the training data to perform parameter tuning via hold-out validation. Note that we use hold-out validation as opposed to cross-validation in order to reduce the risk of overfitting the training data and estimating artificial skill in the verification phase. In the HadISST experiments, SST anomalies are computed by subtracting monthly means of the period 1981--2010. In CCSM4, anomalies are computed relative to the monthly means of the 900 yr training periods. See Eq.~\eqref{eqNino34} ahead for an explicit formula relating SST data vectors to Ni\~no 3.4 indices. 

\subsection{Kernel analog forecasting}
Kernel analog forecasting (KAF)\cite{zhao2016,AlexanderGiannakis19} is a kernel-based nonparametric forecasting technique for partially observed, nonlinear dynamical systems. Specifically, it addresses the problem of predicting the value of a time series $ y( t + \tau ) \in \mathbb{R} $, where $ t $ and $ \tau $ are the forecast initialization and lead times, respectively, given a vector $x(t) \in \mathbb{R}^m $ of predictor variables observed at forecast initialization, under the assumption that $y(t+\tau) $ and $x(t) $ are generated by an underlying dynamical system. To make a prediction of $y(t+\tau)$ at lead time $ \tau = q \, \Delta t$, where $q $ is a non-negative integer and $\Delta t >0$ a fixed interval, it is assumed that a time-ordered dataset consisting of pairs $(x_n, y_{n+q} )$, with 
\begin{equation}
    \label{eqTraining}
    x_n = x( t_n), \quad y_n = y( t_{n+q} ), \quad t_n =  n \, \Delta t, 
\end{equation}
and $ n \in \{ 0, \ldots, N -1  \} $, is available for training. In the examples studied here, $ y( t ) $ is the Ni\~no~3.4 index, $x(t) $ is a lagged sequence of Indo-Pacific SST snapshots (to be defined in Eq.~\eqref{eqDelay} below), $ \Delta t $ is equal to 1 month, and $N$ is the number of samples in the training data. 

For every such lead time $ \tau $, KAF constructs a forecast function $f_ \tau : \mathbb{R}^m \to \mathbb{R}$, such that $ f_\tau( x(t)) $ approximates $y(t+\tau)$. To that end, it treats $x(t) $ and $ y(t) $ as the values of observables (functions of the state) along a trajectory of an abstract dynamical system  (here, the Earth's climate system), operating on a hidden state space $ \Omega$. On $\Omega$, the dynamics is characterized by a flow map $ \Phi^t : \Omega \to \Omega$, such that $ \Phi^t( \omega) $ corresponds to the state reached after dynamical evolution by time $ t $, starting from a state $ \omega \in \Omega$. Moreover, there exist functions $ X : \Omega \to \mathbb{R}^m$ and $Y: \Omega \to \mathbb{R}$ such that, along a dynamical trajectory initialized at an arbitrary state $\omega \in \Omega$, we have $ x(t) = X( \Phi^t( \omega) ) $ and $ y(t) = Y( \Phi^t(\omega)) $. That is, in this picture, $x(t)$ and $y(t)$ are realizations of random variables, $X$ and $Y$, referred to as predictor and response variables, respectively, defined on the state space $\Omega$. 

It is a remarkable fact, first realized in the seminal work of Koopman\cite{Koopman31} and von Neumann\cite{KoopmanVonNeumann32} in the 1930s, that the action of a general nonlinear dynamical system on observables such as $ X $ and $ Y$ can be described in terms of intrinsically linear operators, called Koopman operators. In particular, taking $ \mathcal{F} = \{ f  : \Omega \to \mathbb{R} \} $ to be the vector space of all real-valued observables on $ \Omega $ (note that $Y$ lies in $\mathcal{F}$), the Koopman operator $ U_\tau $ at time $ \tau $ is defined as the linear operator mapping $ f \in \mathcal{F} $ to $ g = U_\tau f \in \mathcal{F} $, such that $ g( \omega ) = f(\Phi^\tau(\omega) ) $. From this perspective, the problem of forecasting $ y(t+\tau ) $ at lead time $ \tau $ becomes a problem of approximating the observable $Y_\tau = U_\tau Y $; for, if $ Y_\tau $ were known, one could compute $   Y_\tau( \Phi^t(\omega) ) = y( t + \tau ) $, where $ \omega \in \Omega $ is the state initializing the observed dynamical trajectory. It should be kept in mind that while $ \mathcal F$ is by construction a linear space, on which $U_\tau $ acts linearly without approximation, the elements of $\mathcal F$ may not  (and in general, will not) be linear functions satisfying a relationship such as  $f(\omega + c \omega') = f( \omega ) + c f(\omega')$ for $ \omega,\omega' \in \Omega $ and $ c \in \mathbb R $. In fact, in many applications $ \Omega $ is not a linear set (e.g., it could be a nonlinear manifold), in which case no elements of $\mathcal F$ are linear functions.

In the setting of forecasting with initial data determined through $X$, a practically useful approximation of $U_\tau Y $  must invariably be through a function $ F_\tau \in \mathcal{F} $ that can be evaluated given values of $ X $ alone. That is, we must have $F_\tau(\Phi^t(\omega)) = f_\tau( x(t) ) $, where $ f_\tau : \mathbb{R}^m \to \mathbb{R} $ is a real-valued function on data space, referred to above as the forecast function, and $ x(t) = X( \Phi^t(\omega ) ) $. In real-world applications, including the ENSO forecasting problem studied here, the observed data $x(t) $ are generally not sufficiently rich to uniquely infer the underlying dynamical state on  $ \Omega $ (i.e., $ X $ is a non-invertible map). In that case, any forecast function $F_\tau $ will generally exhibit a form of irreducible error. The goal then becomes to construct $F_\tau $ optimally given the training data $ ( x_n, y_{n+q}) $ so as to incur minimal error with respect to a suitable metric. Effectively, this is a learning problem for a function in an infinite-dimensional function space, which KAF addresses using kernel methods from statistical learning theory\cite{CuckerSmale01}.    

Following the basic tenets of statistical learning theory, and in particular kernel principal component regression, KAF searches for an optimal $ f_\tau $ in a finite-dimensional \emph{hypothesis space} $ \mathcal{H}_L $, of dimension $L$, which is a subspace of a reproducing kernel Hilbert space (RKHS), $ \mathcal{H}$, of real-valued functions on data space $ \mathbb{R}^m$. For our purposes, the key property that the RKHS structure of $ \mathcal{H}$ allows is that a set of orthonormal basis functions $ \psi_0, \ldots, \psi_{L-1}$ for $\mathcal{H}_L$ can be constructed from the observed data $x_0, \ldots, x_{N-1}$, such that $ \psi_l( x ) $ can be evaluated at any point $ x \in X$, not necessarily lying in the set of training points $x_n$. In particular, $ \psi_l(x(t)) $ can be evaluated for the data $x(t)$ observed at forecast initialization. As with every RKHS, $\mathcal{H}$ is uniquely characterized by a symmetric, positive-definite kernel function $ k : \mathbb{R}^m \times \mathbb{R}^m \to \mathbb{R} $, which can be intuitively thought of as a measure of similarity, or correlation, between data points. As simple example, EOF analysis is based on a covariance kernel, $ k( x, x' ) = x^\top x' $, also known as a linear kernel, but  note that the kernel $ k$ used in our experiments is a Markov-normalized, nonlinear Gaussian kernel (whose construction will be described below).   

Associated with any kernel function $k $ and the training data $ x_0, \ldots, x_{N-1} $ is an $N \times N$ symmetric, positive-semidefinite kernel matrix $\bm K$ with elements $ K_{ij} = k(x_{i-1},x_{j-1}) $, and a corresponding orthonormal basis of $\mathbb{R}^N$ consisting of eigenvectors $  \bm \phi_1, \ldots, \bm \phi_N$, satisfying
\begin{displaymath}
    \bm K \bm \phi_l = \lambda_l \bm \phi_l, \quad \lambda_l \geq 0, \quad \bm \phi_l = ( \phi_{0l}, \ldots, \phi_{N-1,l} )^\top. 
\end{displaymath}
Note that in the case of the covariance kernel, the $ \bm \phi_j $ corresponding to nonzero eigenvalues become principal components given by linear projections of the data onto the corresponding EOF; that is, 
\begin{equation}
    \label{eqLinProj} \bm \phi_l = \bm e_l^\top \bm X / \lambda_l^{1/2},
\end{equation}
where $ \bm X $ is the $ m \times N $ data matrix whose $n$-th column is equal to $ x_{n-1} $, and $ \bm e_l \in \mathbb{R}^m$ is a unit-norm eigenvector (EOF) of the $ m \times m$ spatial covariance matrix $ \bm C = \bm X \bm X^\top$, corresponding to the same eigenvalue $ \lambda_l$,  
\begin{equation}
    \label{eqEOF}
    \bm C \bm e_l = \lambda_l \bm e_l.
\end{equation}
It should also be noted that Eq.~\eqref{eqLinProj} is special to eigenvectors associated with covariance kernels, and in particular does not hold for the nonlinear Gaussian kernels employed in this work.

By convention, we order the eigenvalues $\lambda_l$ in decreasing order. Assuming then that $\lambda_L $ is strictly positive, the basis functions $ \psi_l : \mathbb{R}^m \to \mathbb{R} $ of the hypothesis space $ \mathcal{H}_L$ are given by
\begin{displaymath}
    \psi_l(x) = \frac{1}{N \lambda_l^{1/2}} \sum_{n=0}^{N-1} k( x, x_n ) \phi_{nl}.
\end{displaymath}
Given these basis functions, the KAF forecast function $f_\tau$ at lead time $ \tau = q \, \Delta t$  is expressed as the linear combination 
\begin{displaymath}
    f_\tau = \sum_{l=1}^L\frac{1}{ \lambda_l^{1/2}} c_l(\tau) \psi_l, \quad c_l(\tau) \in \mathbb{R}, 
\end{displaymath}
where the expansion coefficients $ c_l(\tau) $ are determined by regression of the time-shifted response values $ y( t_n + \tau ) = y_{n+q} $ against the eigenvectors $ \bm \phi_l $, viz.  
\begin{displaymath}
    c_l(\tau) = \frac{1}{ N } \sum_{n=0}^{N-1} \phi_{nl} y_{n+q}. 
\end{displaymath}
One can verify that with this choice of expansion coefficients $c_l(\tau)$, $ f_\tau $ minimizes an empirical risk functional $ \mathcal{E}(f_\tau) = \sum_{n=0}^{N} \lvert f_\tau(x_n) - y_{n+q} \rvert^2 / N$ over functions in the hypothesis space $\mathcal{H}_L$.  

A key aspect of the KAF forecast function is its asymptotic behavior as the number of training data $N$ and the dimension of the hypothesis space $L$ grow. In particular, it can be shown\cite{AlexanderGiannakis19} that if the dynamics on $\Omega$ has an ergodic invariant probability measure (i.e., there is a well-defined notion of climatology, which can be sampled from long dynamical trajectories), and the eigenvalues of the kernel matrix $\bm K$ are all strictly positive for every $N$, then under mild regularity assumptions on $X$, $Y$, and the kernel $k$ (related to continuity and finite variance), $ F_\tau $ converges in a joint limit of $N\to \infty$ followed by $L\to\infty$ to the \emph{conditional expectation} $\mathbb{E}(U_\tau Y \mid X)$ of the response observable $U_\tau Y$ evolved under the Koopman operator for the desired lead time, conditioned on the data available at forecast initialization through $X$. In particular, it is a consequence of standard results from probability theory and statistics that $ \mathbb{E}(U_\tau Y \mid X)$ is optimal in the sense of minimal RMSE among all finite-variance approximations of $U_\tau Y $ that depend on the values of $X$ alone. Note that no linearity assumption on the dynamics was made in order to obtain this result. It should also be noted that while, as with many statistical forecasting techniques, stating analogous asymptotic convergence or optimality results in the absence of measure-preserving, ergodic dynamics  is a challenging problem, the KAF formulation described above remains well-posed in quite general settings, including the long-term climate change trends present in the observational SST data studied in this work. 

\subsection{Choice of predictors and kernel} Our choice of predictor function $X$ and  kernel $k$ is guided by two main criteria: (i) $X$ should contain relevant information to ENSO evolution beyond the information present in individual SST snapshots. (ii) $k$ should induce rich hypothesis spaces $\mathcal{H}_L$; in particular, the number of positive eigenvalues $\lambda_l$ (which controls the maximal dimension of $\mathcal{H}_L$) ideally should grow without bound as the size $N$ of the training dataset grows. First, note that the covariance kernel employed in EOF analysis is not suitable from the point of view of the latter criterion, since in that case the number of positive eigenvalues is bounded above by the dimension of the data space $m $ (which also bounds the number of linearly independent, linear functions of the input data). This means that one cannot increase the hypothesis space dimension $L$  to beyond $m$, even if plentiful data is available, i.e., $N \gg m$. In response, following earlier work\cite{GiannakisMajda12a,GiannakisMajda12b,giannakis2015}, to construct $k $ we start from a kernel of the form $ \tilde k(x,x') = h(d(x,x')) $, where $h : \mathbb{R} \to \mathbb{R}$ is a nonlinear shape function, set here to a Gaussian, $ h(u) = e^{- u^2/\epsilon} $ with $\epsilon >0 $, and $ d : \mathbb{R}^m \times \mathbb{R}^m \to \mathbb{R} $ a distance-like function on data space. We set $ d $ to an  anisotropic modification of the Euclidean distance, $ \lVert x - x' \rVert $, shown to improve skill in capturing slow dynamical timescales\cite{giannakis2015}. The kernel $\tilde k $ is then Markov-normalized to obtain $k $ using the normalization procedure introduced in the diffusion maps algorithm\cite{coifman2006diffusion}. With this construction, all eigenvalues of $\bm K $ are positive if the bandwidth parameter $ \epsilon $ is small-enough.    

What remains is to specify the predictor function $X$. For that, we follow the popular approach employed, among many techniques, in singular spectrum analysis (SSA)\cite{GhilEtAl02}, extended EOF analysis, and NLSA\cite{GiannakisMajda12a,GiannakisMajda12b}, which involves augmenting the dimension of data space using the method of delay-coordinate maps\cite{SauerEtAl91}. Specifically, using $ S: \Omega \to \mathbb{R}^p $ to denote the function on state space such that 
\begin{equation}
    \label{eqSST}
    \bm s = S(\omega)
\end{equation}
is equal to the values of the SST field sampled at the $p$ gridpoints on the Indo-Pacific domain corresponding to climate state $ \omega $, we set 
\begin{equation}
    \label{eqDelay}
    X( \omega ) = ( S( \omega ), S( \Phi^{-\Delta t}( \omega ) ), \ldots, S(\Phi^{-(Q-1) \, \Delta t}(\omega))) \in \mathbb{R}^m,    
\end{equation}
where $Q$ is the number of delays, and $ m = pQ $. It is well known from dynamical systems theory\cite{SauerEtAl91} that for sufficiently large $Q$, $ X(\omega) $ generically becomes in one-to-one correspondence with $ \omega $, meaning that as $Q$ grows, $X(\omega)$ becomes a more informative predictor than $S(\omega)$. Elsewhere\cite{Giannakis19,DasGiannakis19}, it was shown that as $Q$ increases, the eigenvectors of the corresponding kernel matrix $\bm K $, and thus the hypothesis spaces $\mathcal{H}_L$, become increasingly adept at capturing intrinsic dynamical timescales associated with the spectrum of the Koopman operator of the dynamical system. In particular, it has been shown\cite{SlawinskaGiannakis17,GiannakisSlawinska18,WangEtAl19} that for interannual embedding windows, $ Q \gtrsim 24$, the leading eigenvectors of $ \bm K $ become highly efficient at capturing coherent modes of variability associated with ENSO evolution, as well as its linkages with the seasonal cycle and decadal variability. Due to this property, the kernels employed in this work are expected to be useful for ENSO prediction since the associated eigenspaces can capture nonlinear functions of the input data, and within those eigenspaces, meaningful representations of ENSO dynamics are possible using a modest number of eigenfunctions.

\subsection{Linear inverse models} 
Under the classical LIM ansatz\cite{penland1993}, the dynamics of ENSO can be well modeled as linear system driven by stochastic external forcing represented as temporally Gaussian white noise. The method used here to conduct LIM experiments follows closely the procedure described by Penland and Sardeshmukh\cite{penland1995}. Specifically, the dynamics is governed by a 
stochastic differential equation
\begin{equation}
\label{LIM}
\dot{\bm \psi}(t) = \bm B \bm \psi(t) + \dot{\bm W}(t),
\end{equation}
where $\bm{\psi}(t) \in \mathbb{R}^L$ is the LIM state vector at time $t$, $\bm{B}$ is an $L \times L$ matrix representing a stable dynamical operator, and $ \bm W(t)$ is a Gaussian white noise process. 

Let $\bm s(t)$ be an SST vector from Eq.~\eqref{eqSST}, observed at forecast initialization time $t$, and $ \bm s'(t) = \bm s(t) - \bar{\bm s}(t) $ be the corresponding anomaly vector relative to the monthly climatology $ \bar{\bm s}(t) $ at time $ t $ (computed as described above for the HadISST and CCSM4 datasets). Then, the corresponding LIM state vector $ \bm \psi(t)$ is given by projection onto the leading $L$ EOFs $ \bm e_l $ from Eq.~\eqref{eqEOF}, computed for the covariance matrix $ \bm C $ based on the anomalies $ \bm s'(t_n) $ in the training data, i.e., 
\begin{displaymath}
    \bm \psi( t ) = (\psi_1(t), \ldots, \psi_L(t))^\top, \quad \psi_l(t) = \bm e_l^\top \bm s'(t) / \lambda_l^{1/2}.
\end{displaymath}
The RMSE-optimal estimate for $\bm{\psi}(t+\tau)$ at an arbitrary lead time $\tau$ under the assumed dynamics in Eq.~\eqref{LIM} is then given by the solution of the deterministic part with initial data $\bm \psi(t)$, viz. 
\begin{displaymath}
    \hat{\bm{\psi}}(t+\tau) =\bm{G} (\tau)\bm{\psi}(t),
\end{displaymath}
where $\bm{G} (\tau) = \exp( \bm{B \tau} ) $. Given this estimate, the predicted Ni\~no~3.4 index at time $ t + \tau $ is given by
    \begin{equation}
        \label{eqLIMPred}
        \hat{y}(t + \tau) = \sum_{l=1}^L z_l \hat \psi_l(t+\tau),
    \end{equation}
    where $ \hat \psi_{l}(t+\tau)$ is the $l$-th component of $\hat{\bm\psi}(t+\tau)$, and $ z_l $ is the regression coefficient of the Ni\~no~3.4 index $y(t_n) $ against the $l$-th SST principal component time series $ \psi_l(t_n) $ in the training phase, 
\begin{displaymath}
    z_l = \frac{1}{N} \sum_{n=0}^{N-1} \psi_l(t_n) y(t_{n}). 
\end{displaymath}
Note that because the Ni\~no 3.4 index $y(t)$ is a linear function of the Indo-Pacific SST anomalies $ \bm s'(t) $, forecasts via Eq.~\eqref{eqLIMPred} are equivalent to (but numerically cheaper than) first computing LIM forecasts $ \hat{\bm s}'(t + \tau)$ of the full Indo-Pacific SST anomalies $ \bm s'( t + \tau ) $, and then deriving from these forecasts a predicted Ni\~no~3.4 index. Specifically, let $J$ denote the set of indices $j$ of the components $s'_j( t ) $ of $ \bm s'( t ) $ lying within the Ni\~no~3.4 region, and $ \lvert J \rvert $ the number of elements of $J$. Then, we have
\begin{equation}
    y(t_n) = \frac{1}{ \lvert J \rvert} \sum_{j \in J} s'_j(t_n),
    \label{eqNino34}
\end{equation}
so that 
\begin{displaymath}
    z_l = \frac{1}{N} \sum_{n=0}^{N-1} \psi_l(t_n) \frac{1}{ \lvert J \rvert} \sum_{j \in J} s'_j(t_n)  = \frac{1}{\lvert J \rvert} \sum_{j\in J} \frac{1}{N} \sum_{n=0}^{N-1} s'_j(t_n) \psi_l(t_n ) = \frac{\lambda_l^{1/2}}{ \lvert J \rvert} \sum_{j\in J}  e_{jl},
\end{displaymath}
where $e_{jl}$ is the $j$-th component of EOF $ \bm e_l $. Noticing that the $j$-th component of the LIM forecast $\hat{\bm s}(t+\tau)$ is given by 
\begin{displaymath}
    \hat s_j( t + \tau ) = \sum_{l=1}^L \lambda_l^{1/2} e_{jl} \hat \psi_l(t + \tau ),
\end{displaymath}
it then follows that
\begin{equation}
    \hat y(t + \tau ) = \sum_{l=1}^L \frac{\lambda_l^{1/2}}{ \lvert J \rvert} \sum_{j\in J} e_{jl} \hat \psi_l(t+\tau) = \frac{1}{ \lvert J \rvert} \sum_{j \in J} \hat s'_j(t + \tau ),  
    \label{eqLIMPred2}
\end{equation}
form which we deduce that predicting using Eq.~\eqref{eqLIMPred} is equivalent to Eq.~\eqref{eqLIMPred2}.

In practice, $\bm{G}(\tau_0)$ is first estimated at some time $\tau_0 = q_0 \, \Delta t$, $ q_0 \in \mathbb{N}$, from the training SST data $\bm{s}(t_n)$ sampled at the times $t_n$ from~\eqref{eqTraining}, 
\begin{equation}
\label{G0}
\bm{G} (\tau_0)=  \left( \frac{1}{N} \sum_{n=0}^{N-1}  \bm{\psi}(t_n+\tau_0) \bm{\psi}(t_n)^\top \right) \left( \frac{1}{N} \sum_{n=0}^{N-1} \bm{\psi}(t_n)\bm{\psi}(t_n)^\top\right)^{-1},
\end{equation}
and then $\bm{G}(\tau)$ is computed at the desired lead time $\tau$ by
\begin{equation}
\label{G}
\bm{G}(\tau) = [\bm{G}(\tau_0)]^{\tau/\tau_0}.
\end{equation}
All LIM-based forecasts reported in this paper were obtained via Eq.~\eqref{eqLIMPred} with $\bm G(\tau)$ from Eq.~\eqref{G}. Note that under analogous ergodicity assumptions to those employed in KAF, the empirical time averages in Eq.~\eqref{G0} converge to climatological ensemble averages. It should also be noted that, among other approximations, the model structure in Eq.~\eqref{LIM} assumes that the dynamics is seasonally independent.

\subsection{Validation}

The main tunable parameters of the KAF method employed here are the number of delays $ Q $, the Gaussian kernel bandwidth $ \epsilon $, and the hypothesis space dimension $L$. Here, we use throughout the values $Q=12 $, $\epsilon = 1$, and $L=400$. As stated above, these values were determined by hold-out validation, i.e., by varying the parameters seeking optimal prediction skill in validation datasets. Note that this search was not particularly exhaustive, as we found fairly mild dependence of forecast skill under modest parameter changes around our nominal values. 

The LIMs in this work have two tunable parameters, namely $\tau_0$ in Eq.~\eqref{G} and the number $L$ of principal components employed. We set $\tau_0=2$ months and $L= 20$, using the same hold-out validation procedure as in KAF.

\subsection{Probabilistic forecasting} Our approach for probabilistic El Ni\~no/La Ni\~na forecasting is based on the standard result from probability theory that the conditional probability of a certain event to occur is equal to the conditional expectation of its associated characteristic function. Specifically, let, as above, $ Y : \Omega \to \mathbb{R} $ be the real-valued function on state space $\Omega$ such that $Y(\omega) $ is equal to the Ni\~no~3.4 index corresponding to climate state $ \omega \in \Omega$. Let also $ \bar Y : \Omega \to \mathbb{R} $ be the 3-month running-averaged Ni\~no~3.4 index, i.e.,
\begin{displaymath}
    \bar Y(\omega) = \frac{1}{3} \sum_{q=-1}^1 Y( \Phi^{q\,\Delta t}(\omega)).
\end{displaymath}
Here, we follow a standard definition for El Ni\~no events\cite{LHeureuxEtAl17}, which declares $\omega$ to be an El Ni\~no state if $ \bar Y(\omega) > 0.5 $ $^\circ $C for a period of five consecutive months about $\omega$. This leads to the characteristic function $ \chi_+ : \Omega \to \mathbb{R}$, such that $ \chi_+( \omega) = 1 $ if there exists a set $ J $ of five consecutive integers in the range $ [ -4, 4 ] $, such that $\bar Y( \Phi^{j \, \Delta t}(\omega) ) > 0.5 $ for all $ j \in J $, and $ \chi_+(\omega) = 0$ otherwise. Similarly, we define a characteristic function $\chi_-$ for La Ni\~na events, requiring that $ \bar Y(\omega) < - 0.5^{\circ}$C for a five-month period about $\omega$. With these definitions, the conditional probabilities $P_{\pm, \tau}(x(t))$ for El Nin\~o/La Ni\~na, respectively, to occur at lead time $\tau $, given the predictor vector $x(t)$ at forecast initialization time $ t $, is equal to the conditional expectation $\mathbb{E}(U_\tau \chi_{\pm} \mid X = x(t))$ of $\chi_{\pm}$ acted upon by the time-$\tau$ Koopman operator $U_\tau$. In particular, because the values of $ \chi_{\pm}$ are available to us over the training period, we can estimate $P_{\pm,\tau}(x(t))$ using the KAF and LIM methodologies analogously to the Ni\~no 3.4 predictions described above. All probabilistic forecast results reported in this paper were obtained in this manner.

\subsection{Forecast skill quantification}
Let $\tilde \omega_n = \Phi^{n \, \Delta t}(\tilde \omega_0)$, with $\tilde\omega_0 \in \Omega$ and $ n \in \{ 0, 1, \ldots, \tilde N-1 \} $, be the climate states in $\Omega$ over a verification period consisting of $\tilde N$ samples, and $ \tilde x_n = X(\tilde \omega_n) $ and $ \tilde y_{n+q} = U_\tau Y(\tilde \omega_n) =  Y( \tilde \omega_{n+q})$ be the corresponding values of the predictor and response (Ni\~no 3.4) functions at lead time $ \tau = q\, \Delta t$. We assess the forecast skill of the prediction function $f_\tau$ for $U_\tau Y$ using the root-mean-square error (RMSE) and Pearson correlation (PC; also known as pattern correlation) scores, defined as 
\begin{displaymath}
    \text{RMSE}(\tau) = \sqrt{\frac{1}{\tilde N}\sum_{n=0}^{\tilde N-1}(f_\tau( \tilde x_n)-\tilde y_{n+q})^2}, \quad \text{PC}(\tau) = \frac{1}{\tilde N}\sum_{n=0}^{\tilde N -1}\frac{(f_\tau(\tilde x_n)-\hat \mu_\tau)( \tilde y_{n+q} - \mu_\tau)}{\hat{\sigma}_\tau\sigma_\tau},
\end{displaymath}
respectively. Here, $\mu_\tau$ ($\hat \mu_\tau$) and $\sigma_\tau$ ($\hat \sigma_\tau$) are the empirical means and standard deviations of $ y_{n+q}$ ($f_\tau(\tilde x_n)$), respectively. 

In the case of the probabilistic El Ni\~no/La Ni\~na forecasts, we additionally employ the BSS and relative-entropy scores, which are known to provide natural metrics for assessing the skill of statistical forecasts\cite{Kleeman02,DelSoleTippett07,WeigelEtAl07,GiannakisEtAl12b}. In what follows, we outline the construction of these scores in the special case of binary response functions, such as the characteristic functions $ \chi_{\pm}$ for El Ni\~no/La Ni\~na events, taking values in the set $  \{ 0, 1 \} $. First, note that every probability measure on $\{ 0, 1\}$ can be characterized by a single real number $\pi \in [ 0, 1 ]$ such that the probability to obtain 0 and 1 is given by $ \pi $ and $1 - \pi $, respectively. Given two such measures characterized by $ \pi, \rho \in [ 0, 1 ] $, we define the quadratic distance function
\begin{displaymath}
    D( \pi, \rho ) = ( \pi - \rho )^2. 
\end{displaymath}
Moreover,  under the condition that $ \rho = 0 $ only if $ \pi = 0 $, we define the relative entropy (Kullback-Leibler divergence)
\begin{displaymath}
    D_\text{KL}(\pi \mid\mid \rho) = \pi \log_2\left( \frac{\pi}{\rho} \right) + (1 - \pi) \log_2\left( \frac{1-\pi}{1-\rho} \right), 
\end{displaymath}
where, by convention, we set $\pi \log_2( \pi / \rho) $ or $(1-\pi)\log_2[ ( 1-\pi )/ (1-\rho)]$ equal to zero whenever $\rho $ or $1-\rho$ is equal to zero, respectively. This quantity has the properties of being non-negative, and vanishing if and only if $\pi = \rho$. Thus, $D_{\text{KL}}$  can be thought of as a distance-like function on probability measures, though note that it is non-symmetric (i.e., in general, $D_{\text{KL}}(\pi \mid\mid\rho) \neq D_{\text{KL}}(\rho \mid\mid \pi)$), and does not obey the triangle inequality. Intuitively, $D_{\text{KL}}( \pi \mid\mid \rho)$ can be interpreted as measuring the precision (additional information content) of the probability measure characterized by $ \pi$ relative to that characterized by $\rho$, or, equivalently, the ignorance (lack of information) of $ \rho$ relative to $\pi$.     

In order to assess the skill of probabilistic El Ni\~no forecasts, for each state $ \tilde \omega_n $ in the verification dataset, we consider three probability measures on $\{ 0, 1 \}$,  characterized by  (i) the predicted probability $P_{+,\tau}(\tilde x_n)$ determined by KAF or LIM; (ii) the climatological probability $ \bar P_+$, which is equal to the fraction of states $\tilde \omega_n$ in the verification dataset corresponding to El Ni\~no events (i.e., those states for which $\chi_+(\tilde \omega_n)=1$); and (iii) the true probability equal to the value $ \chi_+(\tilde \omega_n)$ of the characteristic function indexing the event. Using these probability measures, for each lead time $ \tau = q \, \Delta t $, $ q \in \mathbb{N}_0 $, we define the instantaneous scores 
\begin{gather*}
    \mathcal{B}(\tau;\tilde\omega_n) = D( \chi_+(\tilde \omega_{n+q} ), P_{+,\tau}( \tilde x_n ) ), \quad \bar{\mathcal{B}}(\tau;\tilde\omega_n) = D( \chi_+(\tilde \omega_{n+q}), \bar P_{+}), \\ 
    \mathcal{D}(\tau; \tilde x_n) = D_{\text{KL}}( P_{+,\tau}(\tilde x_n)\mid \mid \bar P_+), \quad \mathcal{E}(\tau; \tilde \omega_n) = D_{\text{KL}}( \chi_+( \tilde \omega_{n+q}) \mid\mid P_{+,\tau}(\tilde x_n) ).
\end{gather*}
Among these, $ \mathcal{B}( \tau; \tilde \omega_n ) $ and $\bar{\mathcal{B}}(\tau;\tilde\omega_n)$ measure the quadratic error of the probabilistic binary forecasts by $ P_{+,\tau}(\tilde x_n) $ and the climatological distribution $ \bar P_+ $, respectively, relative to the true probability $ \chi_+(\omega_{n+q} ) $. In particular, $ \mathcal{B}( \tau; \tilde \omega_n ) $ vanishes if and only if $  P_{+,\tau}(\tilde x_n) $ is equal to 1 whenever  $\chi_{+}(\tilde\omega_{n+q})$ is equal to 1 (i.e., if whenever the forecast model predicts an El Ni\~no event, the event actually occurs). Moreover, based on the interpretation of relative entropy stated above, $ \mathcal{D}(\tau; \tilde x_n)$ measures the additional precision in the forecast distribution relative to climatology, and $ \mathcal{E}(\tau; \tilde \omega_n)$ the ignorance of the forecast distribution relative to the truth. In our assessment of a forecasting framework such as KAF and LIM we consider time-averaged, aggregate scores over the verification dataset, viz.
\begin{gather*}
    \mathcal{B}(\tau) = \frac{1}{\tilde N} \sum_{n=0}^{\tilde N -1 } \mathcal{B}(\tau;\tilde \omega_n), \quad  \bar{\mathcal{B}}(\tau) = \frac{1}{\tilde N} \sum_{n=0}^{\tilde N -1 } \bar{\mathcal{B}}(\tau;\tilde \omega_n), \\
    \mathcal{D}(\tau) = \frac{1}{\tilde N} \sum_{n=0}^{\tilde N-1} \mathcal{D}(\tau; \tilde x_n), \quad \mathcal{E}(\tau) = \frac{1}{\tilde N} \sum_{n=0}^{N-1} \mathcal{E}(\tau; \tilde \omega_n).
\end{gather*}
Note that $\mathcal{B}(\tau)$ is equal to the mean square difference between the characteristic function $ \chi_+ $ and the forecast function $ P_+$ evaluated on the verification dataset. Similarly, $ \bar{\mathcal{B}}(\tau) $ is equal to the mean square difference between $ \chi_+ $ and the constant function equal to the climatological probability $ \bar P_+ $. As is customary, we normalize $\mathcal{B}(\tau) $ by the climatological error $ \bar{\mathcal{B}}(\tau) $, and subtract the result from 1, leading to the Brier skill score\cite{WeigelEtAl07} 
\begin{displaymath}
    \text{BSS}(\tau) = 1 - \frac{\mathcal{B}(\tau)}{\bar{\mathcal{B}}(\tau)}.
\end{displaymath}
Note that in the expression above we have tacitly assumed that the climatological forecast has nonzero error, i.e., $\bar{\mathcal{B}}(\tau)  \neq 0 $, which holds true apart from trivial cases. With these definitions, a skillful model for probabilistic El Ni\~no prediction should have small values of $\text{BSS}(\tau)$, large values of $\mathcal{D}(\tau)$, and small values of $\mathcal{E}(\tau)$.  

Let now $P_{+*} $ be defined such that it is equal to 1 if $ \bar P_+ \leq 0.5 $, and $  0 $ if $ \bar P_+ > 0.5$. Note that the binary  probability distribution represented by $ P_* $ places all probability mass to the outcome in $\{0,1\}$ that is least probable with respect to the climatological distribution, which in the present context corresponds to an El Ni\~no event. One can verify that the precision score $\mathcal{D}(\tau)$ is bounded above by the relative entropy $\mathcal{D}_* = D_{\text{KL}}( P_{+*} \mid \mid \bar P_+)$. The latter quantity provides a natural scale for $\mathcal{D}(\tau)$ corresponding to the precision score of a probabilistic forecast that is maximally precise relative to climatology, in the sense of predicting with probability 1 the climatologically least likely outcome. To define a natural scale for $\mathcal{E}(\tau)$, we consider the time-averaged relative entropy $\mathcal{E}_* = \sum_{n=0}^{\tilde N-1} D_{\text{KL}}( \chi_+(\omega_n) \mid\mid \bar P_+ ) / \tilde N$, i.e., the average ignorance of the climatological distribution relative to the truth. This quantity sets a natural threshold for useful probabilistic El Ni\~no forecasts, in the sense that such forecasts should have $ \mathcal{E}(\tau) < \mathcal{E}_*$. The BSS and relative entropy scores and thresholds for probabilistic La Ni\~na prediction are derived analogously to their El Ni\~no counterparts using the characteristic function $ \chi_{-}$.    

\end{methods}



\bibliographystyle{naturemag}


\begin{addendum}
\item[Acknowledgements] D.~G.\ received support by NSF grants DMS-1521775, 1551489, 1842538, and ONR YIP grant N00014-16-1-2649. X.~W.\ was supported as a PhD student under the first NSF grant. J.~S.\ was supported as a postdoctoral fellow under the second and third NSF grants.  
\item[Author Contributions] X.~W., J.~S., and D.~G.\ designed research. X.~W.\ and J.~S.\ performed numerical experiments. X.~W., J.~S., and D.~G.\ analyzed the results. X.~W.\ and D.~G.\ wrote the main manuscript. All authors reviewed and edited the manuscript.  
 \item[Competing Interests] The authors declare that they have no
competing interests.
\item[Data Availability] The datasets analyzed during the current study are available in the Earth System Grid repository, \url{https: //www.earthsystemgrid.org/dataset/ucar.cgd.ccsm4.joc.b40.1850.track1.1deg.006.html}, and the Haley Centre Ice and Sea Surface Temperature repository, \url{http://www.metoffice.gov.uk/hadobs/hadisst/data/download.html}. The datasets generated during the current study are available from the corresponding author on reasonable request. 
 \item[Correspondence] Correspondence and requests for materials
should be addressed to Dimitrios Giannakis (email: dimitris@cims.nyu.edu).
\end{addendum}


\end{document}